\documentclass[12pt]{iopart}

\usepackage{iopams}  
\expandafter\let\csname equation*\endcsname\relax
\expandafter\let\csname endequation*\endcsname\relax
\usepackage{amsmath}
\usepackage{cases}
\usepackage{amssymb}
\usepackage{color,graphics,epsfig,wasysym}
\usepackage{caption}

\def\be{\begin{equation}}
\def\ee{\end{equation}}
\def\bea{\begin{eqnarray}}
\def\eea{\end{eqnarray}}
\def\bsn{\begin{subnumcases}}
\def\esn{\end{subnumcases}}

\begin{document}

\title[Slow quench dynamics]{Slow quench dynamics in classical systems: kinetic Ising model and zero-range process}
\author{Priyanka$^{1}$, Sayani Chatterjee$^{2,\dagger}$, Kavita Jain$^{2}$}

\address{$^{1}$ Department of Physics (MC 0435) and Center for Soft Matter and Biological Physics, Virginia Tech, Robeson Hall, 850 West Campus Drive, Blacksburg, VA 24061, USA\\
 $^{2}$Theoretical Sciences Unit, 
Jawaharlal Nehru Centre for Advanced Scientific Research, Bangalore 560064, India \\
$^\dagger$ Present address: No. 19 Railway Gate, P. O. Bengal Enamel, North 24 Parganas, West Bengal 743122, India}

\begin{abstract}
While a large number of studies have focused on the nonequilibrium dynamics of a system when it is quenched instantaneously from a disordered phase to an ordered phase, such dynamics have been relatively less explored when the quench occurs at a finite rate. Here we study the slow quench dynamics in two paradigmatic models of classical statistical mechanics, {viz.}, one-dimensional kinetic Ising model and mean-field zero-range process, when the system is annealed slowly to the critical point. Starting from the time evolution equations for the spin-spin correlation function in the Ising model and the mass distribution in the zero-range process, we derive the Kibble-Zurek scaling laws.    
We then test a recent proposal that critical coarsening which is ignored in the Kibble-Zurek argument plays a role in the nonequilibrium dynamics close to the critical point. We find that the defect density in the Ising model and a scaled mass distribution in the zero-range process decay linearly to the respective value at the critical point with the time remaining until the end of the quench provided the final quench point is approached sufficiently fast, and sublinearly otherwise. As the linear scaling for the approach to the critical point also holds when a system following an instantaneous quench is allowed to coarsen for a finite time interval, we conclude that critical coarsening captures the scaling behavior in the vicinity of the critical point if the annealing is not too slow. 
\end{abstract}

\vspace{2pc}

\submitto{\JSTAT}
\maketitle

\clearpage

 \section{Introduction}
 
 The nonequilibrium dynamics following rapid quenches where the annealing time  from a disordered phase to an ordered phase is much shorter than the time scale over which phase ordering occurs have been extensively studied \cite{Bray:1994}. 
In the recent past, attention has focused on slow quenches that are relevant to understand the residual density of defect structures in the slowly cooling early universe \cite{Kibble:1976}, glassy states that are obtained on cooling a liquid at a finite rate  \cite{Cornell:1991,Ritort:2003} and more generally, the nonequilibrium dynamics of classical \cite{Cornell:1992,Biroli:2010,Krapivsky:2010,Jelic:2011,Priyanka:2016,Ricateau:2018,Jeong:2019,Mathey:2020,Sun:2020,Jeong:2020} and quantum systems \cite{Zurek:2005,Polkovnikov:2005,Mukherjee:2007,Dziarmaga:2010,Chandran:2012,Karevski:2016,Zamora:2020} that exhibit second order phase transitions \cite{Zurek:1985}. Theoretical predictions for quantum-annealed systems have also been tested in experiments with different media and especially, ultra-cold atomic gases \cite{Campo:2014,Beugnon:2017}. A general conclusion of these investigations is that there are more defects at the end of the slow quench than in equilibrium. This result may be understood using an argument based on equilibrium state properties \cite{Kibble:1976,Zurek:1985,Zurek:1996}: the idea is that as the correlation length is of order unity in the disordered phase, the system can equilibrate quickly to the changing temperature (or the relevant tuning parameter). However, close to the critical point where the correlation length diverges, the system falls out of equilibrium 
and evolves very slowly resulting in excess defects. Then, assuming that the system does not evolve at all in the nonequilibrium regime, the scaling of the density of excess defects at the end of the quench with the annealing rate can be predicted and has been verified in several studies \cite{Laguna:1997,Yates:1998}. 

The `frozen dynamics' assumption has been questioned recently and using scaling ideas and numerical simulations, it has been argued that the nonequilibrium regime is characterized by critical and sub-critical coarsening which can result in a substantial decrease in the number of excess defects and even change the scaling law from the Kibble-Zurek prediction \cite{Biroli:2010}. 
However, the previous body of work on testing this proposal has focused on the properties at the end of the quench and, to the best of our knowledge, the scaling behavior for the {\it approach} to the final quench point has not been elucidated. 
Here, we study the one-dimensional Ising model with Glauber dynamics for general cooling schemes analytically when it is cooled to zero temperature, and derive scaling properties of the defect density at and close to the end of the quench. We then consider a zero-range process with time-dependent hop rates when it is annealed slowly to the critical point and study such properties analytically and numerically.

It is well known that the classical ferromagnetic Ising model exhibits a nontrivial phase transition above one dimension. The residual defect density on slowly cooling the Ising system from a disordered phase to the critical point or ordered phase has been studied, mainly numerically, in two or higher dimensions without 
\cite{Biroli:2010,Jelic:2011,Liu:2014,Ricateau:2018,Jeong:2019} and with  \cite{Huse:1986,Jackle:1991,Suzuki:2009,Xu:2017} quenched disorder. In Sec.~\ref{Ising}, we study the one-dimensional pure Ising model with Glauber dynamics when it is slowly quenched to zero temperature as it is analytically tractable. Using the known exact solution for the spin-spin correlation function with time-dependent temperature \cite{Reiss:1980,Schilling:1988,Brey:1994}, we calculate the defect density at the end of the quench. Our expression for the dependence of residual defect density on the cooling rate matches with the Kibble-Zurek prediction \cite{Kibble:1976,Zurek:1985,Zurek:1996} as well as the previous work on this  model \cite{Brey:1994,Krapivsky:2010,Jeong:2020}. We also obtain an exact expression for the prefactor of the excess defect density which is in excellent agreement with the recent simulation results \cite{Jeong:2020} and shows that an approximate expression for the prefactor obtained in \cite{Krapivsky:2010} overestimates the defect density. 
In line with the argument of \cite{Biroli:2010}, we find that at the end of the quench, the density of defects is substantially smaller than that expected under `frozen dynamics' assumption.
We then leverage the exact solution to investigate the role of critical coarsening in the approach to zero temperature and find that for sufficiently fast cooling (at a finite rate), the defect density decays with the remaining time until the end of the quench linearly and sublinearly otherwise. As the scaling behavior in the former case is the same as that when the system is quenched instantaneously and allowed to coarsen for a finite time, our main conclusion is that the dynamics in the nonequilibrium regime can be described by the instantaneous quench model, provided the final temperature is approached fast enough. 

In Sec.~\ref{ZRP}, we consider a zero-range process in mean-field geometry and exploit the insights gained from the Ising model to understand its slow quench dynamics. The stationary state of this model is known exactly  in arbitrary dimensions and for hop rates considered here, it exhibits a phase transition from a fluid phase with an order unity particles distributed per site to a condensate phase where a macroscopic number of particles occupy a  site, as the parameter $b$ in the hop rate is increased \cite{Evans:2005}. The instantaneous quench dynamics have been studied using scaling arguments in mean-field geometry and  in one dimension \cite{Godreche:2003}, and the slow quench dynamics in the one-dimensional zero-range process were studied numerically in \cite{Priyanka:2016}. Here, 
we develop a scaling theory for the mean-field zero-range process with time-dependent rates when it is annealed slowly to the critical point and derive the Kibble-Zurek scaling laws for the mass distribution. Our differential equation for the scaling function does not appear to be solvable, and we therefore study the mass distribution close to the critical point numerically and find that, analogous to the Ising model, it varies linearly with the remaining time, provided the parameter $b$ approaches the critical point sufficiently fast.

In Sec.~\ref{disc}, we summarize our results and discuss some open questions. In the following, the quantities pertaining to the stationary state are denoted by calligraphic letters. The nonequilibrium quantities referring to an instantaneous quench are distinguished from those obtained following a slow quench by a caret symbol in the former case.

\section{One-dimensional Ising model with Glauber dynamics}
\label{Ising}

We consider a one-dimensional Ising model on a ring with $L$ sites. In the absence of an external field, the Hamiltonian $H=-J \sum_{i=1}^L \sigma_i \sigma_{i+1}$ where the Ising spin $\sigma_i=\pm 1$ and the coupling $J > 0$. 
Under Glauber dynamics, a spin configuration $\{\sigma_1,...,\sigma_j,...\sigma_L \}$ evolves via single-spin flips, and the distribution of the configuration obeys the following master equation \cite{Glauber:1963}
\be
\frac{\partial \Psi(...,\sigma_j,...)}{\partial t}=\sum_j W(-\sigma_j \to \sigma_j)  {\Psi}(...,-\sigma_j,...) -W(\sigma_j \to -\sigma_j)  {\Psi}(...,\sigma_j,...)~,
\ee
where the transition rate 
\be
W( \sigma_j \to -\sigma_j)=\frac{A}{2}  \left( 1-\frac{\gamma}{2} \sigma_j (\sigma_{j-1}+\sigma_{j+1}) \right)~,
\label{Irate}
\ee
and 
\bea
\gamma(T) &=& \tanh \left( \frac{2 J}{k_B T} \right)~,\\
A(T) &=& A_0 \exp  \left(-\frac{\Delta}{k_B T} \right)~.
\eea
As the parameter $\gamma$ decreases with decreasing temperature $T$, by virtue of (\ref{Irate}), the spins tend to align with each other at low temperatures. 
The Arrhenius rate $A$ introduces an activation energy $\Delta$ for a spin flip to occur and is evidently  important at low temperatures \cite{Reiss:1980,Brey:1994}.

Motivated by the glass-problem in which the dynamics are `frozen'  in a slowly cooled liquid at low temperatures due to the activation barrier \cite{Ritort:2003}, the slow quench dynamics in the one-dimensional Ising model have been studied analytically by including a time-dependence in the temperature \cite{Reiss:1980,Schilling:1988,Jackle:1991,Brey:1994}. However, even in the absence of an activation barrier, the dynamics slow down at low temperatures as the relaxation time becomes longer than the annealing time \cite{Kibble:1976,Zurek:1985,Zurek:1996}; here, we are interested in understanding such dynamics and therefore, we will set $\Delta=0$ throughout the following discussion. For convenience, we will also set $J/k_B=1$. 

\subsection{General expression for the two-point correlation function}

We consider the two-point correlation function $G_k(t)=\langle \sigma_i \sigma_{i+k} \rangle$ where the angular brackets denote an average with respect to the ensemble distribution ${\Psi}(\vec \sigma, t)$. For arbitrary $\gamma$, it obeys the following exact equation \cite{Glauber:1963,Reiss:1980}, 
\bea
{\dot G_k} =-2 G_k+ \gamma(t) (G_{k-1}+G_{k+1})~,~k=1,...,L-1~,
\label{fullGeqn}
\eea
where the dot denotes a time derivative and the boundary condition $G_{0}(t)=G_L(t)=1$ (for a generalization to higher-point correlation function, see \cite{Aliev:2009}). 

When the temperature is constant in time, the dynamics of the two-point correlation function have been studied in detail \cite{Glauber:1963} (see Cornell in \cite{Privman:2005} for a review). For an infinitely large system, at large times, the equilibrium correlation function is given by 
\be
G_k(t\to \infty) \equiv {\cal G}_k= \left(\frac{1-\sqrt{1-\gamma^2}}{\gamma} \right)^k~,~k=0, 1,....
\label{Gequil}
\ee
At low temperatures ($T \to 0$), ${\cal G}_k \approx e^{-k/\xi}$ where the correlation length $\xi=[2 (1-\gamma)]^{-1/2} \sim e^{2/T}$ which  diverges as the temperature approaches zero. As a result, the associated time scale also diverges; in particular, the local magnetization $\langle \sigma_i(t) \rangle$ in equilibrium relaxes in a time $\propto \xi^{z}$ with the dynamic critical exponent $z=2$. If the system is instantaneously quenched from high temperature to a low temperature (including $T=0$), the equal-time correlation function ${\hat {G}}_k(t)={\hat {F}}(k/\ell(t))$ where the domain length $\ell(t) \sim t^{1/{\hat z}}$ and the coarsening exponent ${\hat z}=2$.

We are interested in understanding the dynamics when the Ising system initially in an equilibrium state at infinite temperature is cooled to zero temperature in finite time $\tau$ (the assumption of infinite $T(0)$ is not especially restrictive but simplifies the analysis). For time-dependent $\gamma$, an explicit expression for the two-point correlation function has been obtained in \cite{Reiss:1980} and \cite{Brey:1994}. Here, we follow the latter where it has been shown that for an infinitely large system, the deviation of the correlation function $G_k(t)$ at time $t$ from the corresponding equilibrium value at instantaneous temperature is given by  \cite{Brey:1994}
\bea
{\cal G}_k(t)&-&G_k(t) =\int_0^\pi dq \phi_k(q)  \int_0^t dt' e^{-2 \int_{t'}^t dy (1-\gamma(y) \cos q)} \sum_{m=1}^\infty \phi_m(q) {\dot {\cal G}}_m(t')\nonumber \\
&+& \int_0^\pi dq \phi_k(q) e^{-2 \int_{0}^t dy (1-\gamma(y) \cos q)} \sum_{m=1}^\infty \phi_m(q) ({\cal G}_m(0)-G_m(0)) ~,
\label{Iexact}
\eea
where $\phi_k(q)=\sqrt{\frac{2}{\pi}} \sin(k q)$ obeys the eigenvalue equation $\gamma (\phi_{k-1}(q)+\phi_{k-1}(q))-2 \phi_k(q)=-\lambda \phi_k(q)$ with eigenvalue $\lambda(q)=2 (1-\gamma \cos (q))$. The above expression is obtained on expanding ${\cal G}_k(t)-G_k(t)$ as a linear combination of the eigenfunctions $\phi_k(q)$ and using the time evolution equation (\ref{fullGeqn}). 
Furthermore, since the system is assumed to be initially in the equilibrium state at infinite temperature, $G_k(0)={\cal G}_k(0)=\delta_{k,0}$; on carrying out the sum over $m$ in the first term on the right-hand side (RHS) of (\ref{Iexact}) exactly, we obtain
\bea
\hspace{-0.3in}{\cal G}_k(t)-G_k(t)=\int_0^t dt' e^{-2 (t-t')} {\dot \gamma(t')} \int_0^\pi \frac{dq}{\pi}   \frac{ \sin(q) \sin(k q)}{(1-\gamma(t') \cos(q))^2} e^{2 \cos q \int_{t'}^t dy \gamma(y)} ~.
\label{Ispl}
\eea
In the following subsections, we will analyze the above double integral in detail. 

Following previous work \cite{Brey:1994,Krapivsky:2010}, we consider the following class of cooling protocols, 
\begin{subnumcases}{\gamma(t,\tau) =}
1- \left(1- x \right)^\alpha&,~$\alpha > 0$ \label{Gpower} \\
1- \exp \left[1- \left(1-x \right)^{-\beta} \right] &,~$\beta > 0$ \label{Gexpo}~,
\end{subnumcases}
where $x=t/\tau$. The parameter $\gamma(x)$ increases monotonically from zero to one with increasing time and, (\ref{Gpower}) and (\ref{Gexpo}), respectively, correspond to temperature, 
\begin{subnumcases}{T(x) \stackrel{t \to \tau}{\sim}}
\frac{-4}{\alpha \ln (1- x)}  &\textrm{(logarithmic cooling)}\\
4 \left(1-x \right)^{\beta}  &\textrm{(algebraic cooling)}~.
\end{subnumcases}
Note that the above cooling schemes reduce to the instantaneous quench limit when $\alpha, \beta \to  \infty$.

\subsection{Dynamics at high temperatures}

\begin{figure}[t]
\begin{center}
\includegraphics[width=12cm]{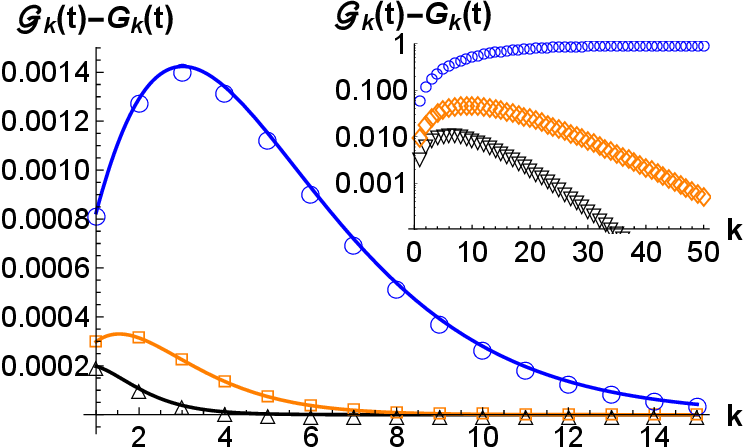}
\caption{Kinetic Ising model: Deviation in the correlation function, ${\cal G}_k(t)-G_k(t)$ for logarithmic cooling scheme (\ref{Gpower}) with $\tau=2 \times 10^3$ and $\alpha=3/2$ at $t=500 (\vartriangle), 1000 (\square), 1500 (\Circle), 1800 (\triangledown), 1900 (\lozenge), 2000 (\circ)$ that corresponds to the temperature $5.47, 2.6, 1.48, 0.97, 0.77, 0$, respectively. These data are obtained by numerically solving the dynamical equation (\ref{fullGeqn}) and using the equilibrium correlation function (\ref{Gequil}). The lines show the expression (\ref{devpara}) for $t \leq 1500$ in the main figure. The deviation ${\cal G}_k(t)-G_k(t)$   increases with time, and is substantial when the system is close to zero temperature (see inset).}
\label{fig_coraw}
\end{center}
\end{figure}

We first consider the parameter regime where $\gamma \ll 1$. Since the integrand in the outer integral of (\ref{Ispl}) suppresses large $t-t'$, the inner integral receives most contribution when $\gamma$ is evaluated in the vicinity of $t$ thus yielding 
\bea
 {\cal G}_k(t)-G_k(t)&\approx& \int_0^t dt' \int_0^\pi \frac{dq}{\pi}  \frac{\dot \gamma(t) e^{-2 (t-t') (1- \gamma(t) \cos(q))} \sin (q) \sin(k q)}{(1-\gamma(t) \cos (q))^2} \\
&=& \frac{{\dot \gamma(t)}}{2 \pi} \int_0^{\pi} dq \frac{\sin(k q) \sin (q)}{(1- \gamma(t) \cos (q))^3} \\
&=& \frac{k}{4} \frac{{\dot \gamma}}{\gamma} \frac{{\cal G}_k(t)}{1-\gamma^2} \left(k+\frac{1}{\sqrt{1-\gamma^2}} \right)~.
\label{devpara}
\eea

At short times, as Fig.~\ref{fig_coraw} shows, the correlation function ${G}_k(t)$ is close to the equilibrium correlation function ${\cal G}_k(T(t))$ at instantaneous temperature. But, with increasing time, the deviation increases as the system relaxes at a rate slower than the rate of change in the temperature. For a given $t \ll \tau$, the deviation ${\cal G}_k(t)-G_k(t) \propto k^2 x^k$ is, however, nonmonotonic in the inter-spin distance with the peak occurring at larger $k$ for larger times.  Figure~\ref{fig_coraw} also shows that the nonequilibrium correlation function, $G_k(t)$ is always smaller than the equilibrium correlation function, ${\cal G}_k(t)$; this is because the nonequilibrium system has not completely relaxed to the  stationary state at the instantaneous temperature and is effectively at a slightly higher temperature.

\subsection{Dynamics at low temperatures}

Close to zero temperature, the behavior of the correlation function is determined by how the function $\gamma$ approaches one. We therefore consider the logarithmic and algebraic cooling schemes separately. 
\subsubsection{Logarithmic cooling}

Using the expression (\ref{Gpower}) for $\gamma$ in (\ref{Ispl}), we obtain
\be
{\cal G}_k(t)- G_k(t)=\frac{\alpha}{\pi} \int_0^x dy (1-x+y)^{\alpha-1} \int_0^\pi dq \frac{\sin(q) \sin(k q) e^{-2 \tau z(x,y,q)}}{\left[1-\cos(q)+(1-x+y)^\alpha \cos(q) \right]^2}~,
\label{powint1}
\ee
where 
\be
z=(1-\cos(q)) y+\left[\frac{(1-x+y)^{\alpha+1}-(1-x)^{\alpha+1}}{\alpha+1} \right]\cos(q) ~,
\label{smallq}
\ee
and $x=t/\tau$. For $\tau \gg 1$ and $x \to 1$, the exponential term in the inner integral on the RHS of (\ref{powint1}) contributes the most when $q$ and $y$ are close to zero. Therefore, for large $\tau$, we obtain 
\be
{\cal G}_k(t)-G_k(t) \approx \frac{\alpha}{\pi} \int_0^x dy (1-x+y)^{\alpha-1} \int_0^\infty dq \frac{q \sin(k q) e^{-2 \tau \left[\frac{y q^2}{2}+\frac{(1-x+y)^{\alpha+1}-(1-x)^{\alpha+1}}{\alpha+1} \right]}}{\left[\frac{q^2}{2}+(1-x+y)^\alpha \right]^2}~.
\label{sine}
\ee
Then a change of variables to $Y=y \tau^{\frac{1}{1+\alpha}}$ and  $Q=q \tau^{\frac{\alpha}{2(1+\alpha)}}$ immediately shows that the deviation in the correlation function has a scaling form where the scaling function is given by
\bea
{\cal G}_k(t)-G_k(t) \equiv F(K,Z) \nonumber \\
\hspace{-0.2in}=\frac{\alpha}{\pi} \int_0^\infty dY  (Z+Y)^{\alpha-1} e^{-\frac{2}{\alpha+1} \left[(Z+Y)^{\alpha+1}-Z^{\alpha+1} \right]}   \int_0^\infty dQ \frac{Q \sin(K Q) e^{-Y Q^2}}{\left[\frac{Q^2}{2}+(Z+Y)^\alpha \right]^2}
\label{gpowCk} 
\eea
and the scaling variables are 
\be
K=k \tau^{-\frac{\alpha}{2 (1+\alpha)}}, Z=\frac{1-x}{1-{x_*}}=(1-x) \tau^{\frac{1}{1+\alpha}}~.
\label{KZexpo1}
\ee

\begin{figure}[t]
\begin{center}
\includegraphics[width=12cm]{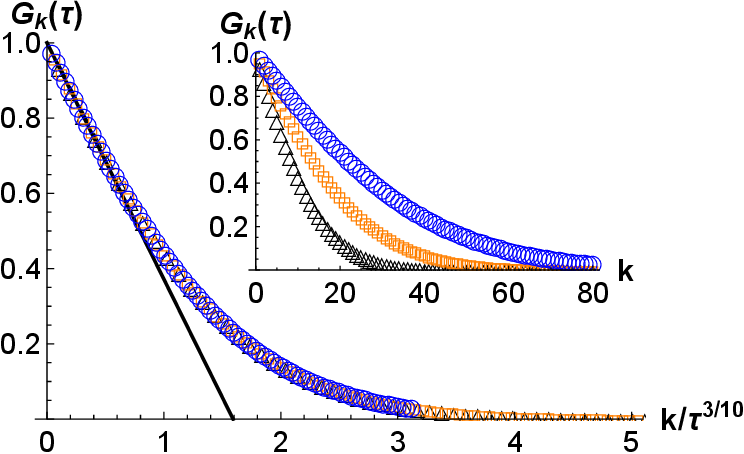}
\caption{Kinetic Ising model: Nonequilibrium correlation function, $G_k(\tau)$ at zero temperature for logarithmic cooling scheme (\ref{Gpower}) and $\alpha=3/2$ with $\tau=2 \times 10^3 (\vartriangle), 10^4 (\square), 5 \times 10^4 (\circ)$. The points show the data obtained by numerically solving the dynamical equation (\ref{fullGeqn}), and the line shows the expression (\ref{powCP1}) for small $k$.}
\label{fig_corrcp}
\end{center}
\end{figure}

In a numerical study of this model \cite{Jeong:2020}, the deviation ${\cal G}_k(t)-G_k(t)$ was found to be a nonmonotonic function of time with the peak time scaling as $x_*$ in accordance with the above scaling form. Note that at low temperatures ($T \to 0$), since the equilibrium correlation function given by (\ref{Gequil}) can be written as
\be
{\cal G}_k(t) \approx e^{-k \sqrt{2 (1-\gamma)}}=e^{-\sqrt{2} K Z^{\alpha/2}}, 
\label{equilsc}
\ee
the nonequilibrium correlation function $G_k(t)$ is also a function of $K$ and $Z$, as also confirmed by the data collapse in Fig.~\ref{fig_corrcp} for $G_k(\tau)$.

The scaling exponents in (\ref{KZexpo1}) are in agreement with the Kibble-Zurek argument \cite{Kibble:1976,Zurek:1985,Zurek:1996} which states that 
for $x < {x_*}$, the system stays close to the equilibrium state at the instantaneous temperature. But, for $x > {x_*}$, it falls out of equilibrium as the equilibrium relaxation time, $\xi^z$ exceeds the time remaining to reach the final quench temperature. Therefore, at $x=x_*$ where these two time scales are of the same order, we have $\tau ( 1-{ x_*}) \sim \xi^z(x_*)$ where the correlation length $\xi(x)= 2^{-1/2}(1-x)^{-\alpha/2}$ for logarithmic cooling (see (\ref{equilsc})) and the dynamic exponent $z=2$ \cite{Glauber:1963}. This immediately yields the Kibble-Zurek time and length scale, $1-{x_*} \sim \tau^{-\frac{1}{1+\alpha}}$ and $\xi(x_*) \sim \tau^{\frac{\alpha}{2 (1+\alpha)}}$, respectively.

The double integral in (\ref{gpowCk}) does not appear to be exactly solvable. But one can find the scaling function $F(K, Z)$ in various limits of interest.  We first consider the deviation in the correlation function at the end of the quench ($Z=0$), and find that 
\bsn
{1- G_k(\tau)=F(K,0)=}
 \frac{K}{\sqrt{\pi}} \left(\frac{2}{1+\alpha} \right)^{\frac{1}{2(1+\alpha)}} \Gamma \left( \frac{1+2 \alpha}{2+2 \alpha}\right) &,~$K \ll 1$ \label{powCP1}\\
1 &,~ $K \gg 1$ ~. \label{powCP2}
\esn
For finite $k$, on expanding the sine function in (\ref{sine}) for small argument and performing the resulting integrals, we obtain (\ref{powCP1}). For large $K$, our numerical study of (\ref{sine}) suggests that the correlations decay to zero as a stretched exponential.

\begin{figure}
\begin{center}
\includegraphics[width=12cm]{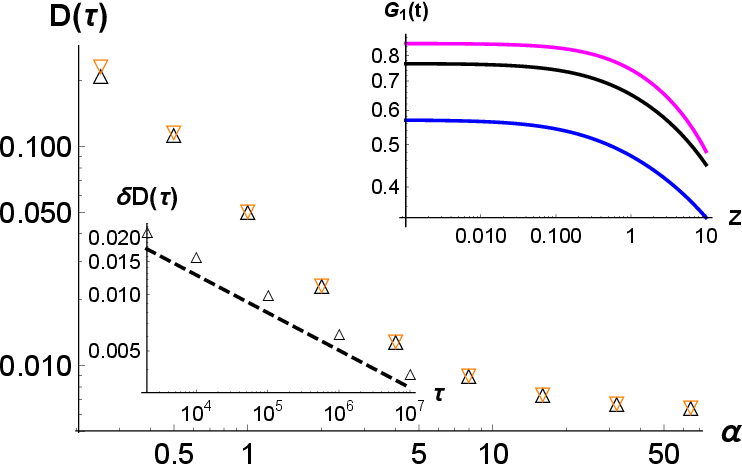}
\caption{Kinetic Ising model: Defect density $D(\tau)$ at zero temperature for logarithmic cooling scheme (\ref{Gpower}) and $\tau=2000$ as a function of the quench exponent $\alpha$. The numerical data  ($\vartriangle$)  does not agree well with the analytical result (\ref{powCP1}) ($\triangledown$) for small $\alpha$ since the correction $\delta D(\tau)$ to the leading order result decays slowly (see bottom, left inset for $\alpha=1/4$). The top, right inset shows the correlation function $G_1(t)$ for $\tau=2000$ and $\alpha=1/4,1/2,3/4$ (bottom to top) as a function of the scaled time $Z=(1-x) \tau^{\frac{1}{1+\alpha}}$. Note that $G_1(t)$ changes very slowly for $Z < 1$.}
\label{fig_defden}
\end{center}
\end{figure}

The domain wall density, ${D}(\tau)=(1-G_1(\tau))/2$ at the end of the quench has been previously studied analytically \cite{Krapivsky:2010} and numerically \cite{Jeong:2020}. 
Although the approximate expression (13) in \cite{Krapivsky:2010} (with $A=1$) captures the $\tau$-dependence of the defect density correctly, it overestimates the prefactor for $\alpha > 0$. The ratio of Krapivsky's estimate of the prefactor to our corresponding exact result obtained using (\ref{powCP1}) is equal to $[\frac{2}{\pi} \Gamma \left(1-\frac{1}{2+2 \alpha}\right) \Gamma \left(1+\frac{1}{2+2 \alpha}\right)]^{-1}$ which 
 increases with increasing $\alpha$. For $\alpha=1$ and $2$, this ratio is equal to $\sqrt{2}$ and $3/2$, respectively, in agreement with the numerical estimates of this discrepancy in \cite{Jeong:2020}. Figure~\ref{fig_defden} shows that our (\ref{powCP1}) for large $\tau$ is in excellent agreement with the numerical data except for small $\alpha$; this is because the subleading correction to (\ref{powCP1}) scales as $\tau^{-\alpha/(1+\alpha)}$, as can be seen by retaining the next order term in (\ref{smallq}) and using (\ref{KZexpo1}). 
 We also verify that in the instantaneous quench limit $\alpha \to \infty$, (\ref{powCP1}) reduces to (\ref{rapid2}) at $Z=0$ (see below).

Equation (\ref{powCP1}) shows that the defect density at zero temperature decreases with increasing $\tau$, as intuitively expected.  
The defect density $D(\tau)$ also decreases with increasing $\alpha$ which can be understood if we reject the `frozen dynamics' assumption \cite{Laguna:1997,Yates:1998}: for large $\alpha$, the system reaches low temperature quickly and spends the remaining time until the end of the quench (given by (\ref{KZexpo1})) equilibrating thereby decreasing the number of defects.


\subsubsection{Algebraic cooling}

Here, we focus on the small-$k$ behavior. Using the cooling scheme (\ref{Gexpo}) in (\ref{Ispl}), we find that
\bea
{\cal G}_k(t)-G_k(t) 
&\approx& \frac{k \beta}{\pi} \int_0^x dy e^{1-(1-x+y)^{-\beta}} (1-x+y)^{-1-\beta} \nonumber \\
&\times& \int_0^\pi dq \frac{q \sin(q) e^{-2 \tau z(x,y,q)}}{[1+(e^{1-(1-x+y)^{-\beta}}-1) \cos(q)]^2} ~,
\eea
where $z=y (1-\cos(q))+\cos(q) \int_{x-y}^x dy' e^{1-(1-y')^{-\beta}}$. As in the last subsection, we expand the integrand in the inner integral for small $q$. Then, for large $\tau$ and $x \to 1$, we  obtain
\bea
 {\cal G}_k(t)-G_k(t) &\approx& 
 \frac{k}{\pi \sqrt{\tau}} \int_0^\infty \frac{dW}{\sqrt{y(W)}} \int_0^\infty dQ \frac{Q^2 e^{-Q^2}}{(\frac{Q^2}{2}+W)^2}~,
\eea
where $W=\tau y e^{1-(1-x+y)^{-\beta}}$ is finite when $x \to 1, y \to 0, \tau \to \infty$. This implies that $1-x+y \approx (\ln \tau)^{-1/\beta}$ using which we find that
\bsn
{{\cal G}_k(t)-G_k(t)=}
\frac{K}{\sqrt{\pi(1-Z)}} ~&,~ $Z < 1$ \label{expCP1}\\
0 ~&,~ $Z > 1$
\esn
where 
\be
K=k \sqrt{\frac{(\ln \tau)^{1/\beta}}{\tau}}~,~Z=(1-x) (\ln \tau)^{1/\beta}~.
\label{KZexpo}
\ee

Equation (\ref{KZexpo}) is in agreement with the Kibble-Zurek argument \cite{Kibble:1976,Zurek:1985,Zurek:1996}: as for logarithmic cooling, on equating the relaxation time $\sim e^{-(1-x_*)^{-\beta}}$ and the remaining time $\tau (1-x_*)$, for large $\tau$, one obtains $1-x_* \sim (\ln \tau)^{-1/\beta}$ and  $\xi(x_*) \sim \tau (\ln \tau)^{-1/\beta}$. Equation (\ref{expCP1}) shows that the deviation in the correlation function, when expressed in terms of the scaling variables, is independent of the quench exponent $\beta$ and equals (\ref{rapid2}) at $Z=0$ which, as mentioned above, is obtained when $\alpha \to \infty$; this result is a simple consequence of the fact that for $x \to 1$, the function $\gamma(x)$ in (\ref{Gexpo})  decays faster than any power law. On comparing (\ref{expCP1}) with (15) of \cite{Krapivsky:2010}, we find that the prefactor in the expression of the defect density is overestimated by a factor $\pi$ in \cite{Krapivsky:2010}.

\subsection{Relation to the coarsening process}

The correlation function $G_1(t)$ displayed in the inset of Fig.~\ref{fig_defden} for logarithmic cooling decays fast with time for $Z \gg 1$ but it changes very slowly (``frozen") at low temperatures.  
In view of this behavior, the Kibble-Zurek argument assumes that the dynamics remain frozen for times larger than ${t_*}=\tau x_*$,  and the density of defects at the end of the quench is simply inherited from that at $t={t_*}$ \cite{Kibble:1976,Zurek:1985,Zurek:1996}. 
However, from (\ref{equilsc}) and (\ref{powCP1}), we find that 
\be
\frac{D(\tau)}{{\cal D}({t_*})} =\frac{1-G_1(\tau)}{1-{\cal G}_1({t_*})} < 1 ~,~ \forall ~\alpha > 0~.
\ee
For logarithmic cooling, the above ratio is approximately $0.59, 0.49, 0.38$ for $\alpha=1/2, 1, 10$, respectively, which shows that the defect density is substantially smaller than that predicted by the Kibble-Zurek argument (except, of course, for $\alpha=0$ for which the remaining time is of order one). It has been proposed that during the time interval ${t_*} < t < \tau$, the dynamics are not frozen and critical coarsening takes place that results in the decrease in the number of domain walls and can even change the scaling law as the dynamic correlation length at the end of the quench scales as $(\tau-t_*)^{1/{\hat z}}$ \cite{Biroli:2010}.   
 
To investigate this proposal, we first note that if the system is instantaneously quenched from an initial temperature $T_i=T({t_*})$ to a final temperature $T_f=T(\tau)$, the nonequilibrium correlation function following an instantaneous quench is given by \cite{Glauber:1963}, 
\bea
{\hat {G}}_k(t) &=&y_f^k+\frac{2}{\pi}  \int_0^\pi dq e^{-2 (1-\gamma \cos(q)) (t-{t_*})} \sin(k q) \nonumber \\
&\times & \left[ \frac{y_i \sin(q)}{1-2 y_i \cos(q)+y_i^2} -\frac{y_f \sin(q)}{1-2 y_f \cos(q)+y_f^2} \right]  ~,~t > {t_*}~,
\eea
where $y_{i,f}=\tanh(T_{i,f}^{-1})$. As the above integral gets a dominant contribution from small $q$ and  $y_f=1$, the first term in the bracket on the RHS of the above equation can be neglected in comparison to the second term. On carrying out the resulting integral, we find that 
\bea
{\hat {G}}_k(t) \approx 1-\frac{k}{\sqrt{\pi (t-{t_*})}}  ~,~~ k \ll \sqrt{t-{t_*}}~.
\label{rapid1}
\eea
Using $K=\frac{k}{\sqrt{\tau (1-x_*)}}, Z=\frac{1-x}{1-x_*}$, the above equation can be rewritten as
\be
{\hat {G}}_k(t)=1-\frac{K}{\sqrt{\pi}}- \frac{K Z}{2\sqrt{\pi}} ~,~ K, Z \ll 1~.
\label{rapid2}
\ee

Now, assuming that the critical coarsening drives the dynamics for $t > {t_*}$  \cite{Biroli:2010},  $G_k(t)$ may be approximated by ${\hat {G}}_k(t)$.  The above equation then shows that the density of defects at the end of the quench has the same $\tau$-scaling as one would obtain by assuming that the dynamics remain frozen; however, this conclusion is a consequence of the fact that the exponent $z={\hat z}=2$ in this model. But, using the above results, we also find that the ratio of the density of defects, $(1-G_1(\tau))/(1-{\hat {G}}_1(\tau))$, in general, differs from one.    
Importantly, (\ref{rapid2}) shows that at low temperatures, the correlation function ${\hat {G}}_k(t)$ depends {\it linearly} on $Z$ for any cooling protocol. Therefore, below we study the scaling behavior of $G_1(t)$ close to zero temperature in order to test how well the instantaneous quench model describes the dynamics when ${t_*} < t < \tau$. 

\subsubsection{Logarithmic cooling}


For small $K$ and nonzero $Z$, from (\ref{gpowCk}), we obtain
\bea
{\cal G}_k(t) &-&G_k(t)= \alpha K \int_0^\infty dY  (Z+Y)^{\alpha-1} e^{-\frac{2}{\alpha+1} \left[(Z+Y)^{\alpha+1}-Z^{\alpha+1} \right]} \nonumber \\
&\times& \left[   \frac{1+4 Y (Z+Y)^\alpha}{\sqrt{2 (Z+Y)^\alpha}} e^{2 Y (Z+Y)^\alpha} \textrm{erfc}(\sqrt{2 Y (Z+Y)^\alpha}) -2 \sqrt{\frac{Y}{\pi}}\right] ~. \label{smallz1}
\eea
On changing the dummy variable from $Y$ to $w=\frac{2}{\alpha+1}(Z+Y)^{\alpha+1}$ in the above integral and denoting the resulting integrand by $e^{-w} H(Z,w)$, we obtain
\bea
{\cal G}_k(t) -G_k(t) &=& K \alpha \int_{\frac{2 Z^{\alpha+1}}{\alpha+1}}^\infty dw e^{-w} H(Z,w) \\
&\stackrel{Z \to 0}{\approx}& K \alpha \int_{\frac{2 Z^{\alpha+1}}{\alpha+1}}^\infty dw e^{-w} \left( H(0,w)+Z \frac{\partial H(0,w)}{\partial w} \right) ~,\label{Heqn}
\eea
where the subleading terms in the above integrand are of the order $Z^{1+\alpha}$ and $Z^2$ for $\alpha \leq 1$ and $\alpha >1$, respectively. As the integral over the first term on the RHS of (\ref{Heqn}) is equal to $1-G_k(\tau)-\sqrt{2} K Z^{\alpha/2}$, we find that to leading order in $Z$, 
\bsn
{G_k(\tau)-G_k(t)=} 
\frac{K \alpha Z}{\pi} \int_{\frac{2 Z^{\alpha+1}}{\alpha+1}}^\infty dw \left(\frac{2}{(\alpha+1) w} \right)^{\frac{3}{2 (\alpha+1)}}{I}(w) ~&,~ $\alpha <1/2$ \label{integ} \\
\frac{K \alpha Z}{\pi} \int_{0}^\infty dw \left(\frac{2}{(\alpha+1) w} \right)^{\frac{3}{2 (\alpha+1)}} {I}(w) ~&,~ $\alpha > 1/2$
\esn
where
\be
I(w)=\sqrt{\pi} e^{-w} (\alpha
   w+w+1)-\frac{\pi}{2}  e^{\alpha w}  (2 (\alpha+1) w+3) \sqrt{(\alpha+1) w}\text{erfc}\left(\sqrt{(\alpha+1)
   w}\right)~.
\ee
On performing the above integrals, we finally arrive at
\bsn
{G_k(\tau)-G_k(t) \stackrel{Z \ll 1}{=}} 
\frac{4 K \alpha}{\sqrt{\pi} (1-2 \alpha)} Z^{\alpha+\frac{1}{2}} ~&,~ $\alpha <1/2$ \label{slowzs1}\\
\frac{2K \alpha}{\sqrt{\pi}} \ln(1/Z) Z~&,~ $\alpha=1/2$\\
\frac{K }{2  \sqrt{\pi}} \left(\frac{2}{1+\alpha}\right)^{\frac{3}{2 (1+\alpha)}} \Gamma\left( \frac{2 \alpha-1}{2 \alpha+2} \right) Z~&,~$\alpha > 1/2$~. \label{slowzs3}
\esn

\begin{figure}[t]
\begin{center}
\includegraphics[width=12cm]{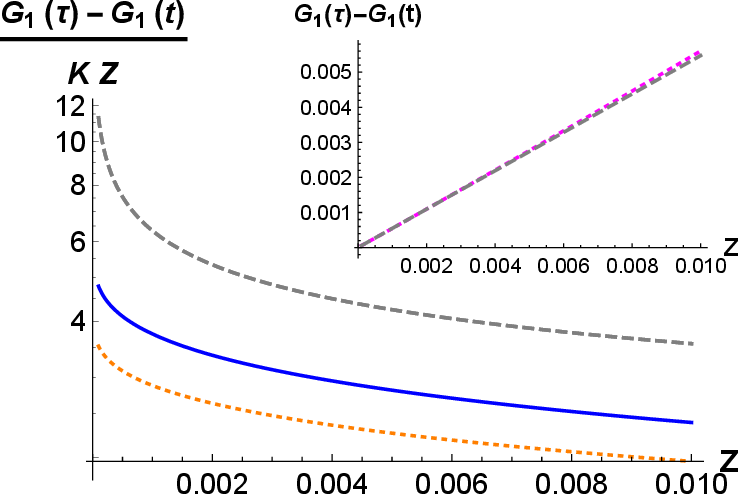}
\caption{Kinetic Ising model: (Scaled) correlation function ${G}_1(\tau)-G_1(t)$ at low temperatures as a function of  $Z=(1-x) \tau^{\frac{1}{1+\alpha}}$ for logarithmic cooling scheme (\ref{Gpower}) and $\alpha=1/4$ with $\tau=10^8$ (dotted) and $10^{10}$ (solid) obtained by solving (\ref{fullGeqn}) numerically. The topmost dashed curve shows the analytical result (\ref{slowzs1}). As the curves diverge for $Z \to 0$, it follows that the correlation function $G_1(t)$ decays sublinearly with $Z$ for $\alpha < 1/2$. In contrast, the nearest-neighbor correlation exhibits linear decay for $\alpha >  1/2$, as the inset shows for $\alpha=3/2$ with $\tau=10^4$; here, the numerical data is shown in dots and the analytical expression (\ref{slowzs3}) by dashed line.}
\label{fig_coarsen}
\end{center}
\end{figure}

We first verify that the result (\ref{slowzs3}) matches the numerical data for finite $\alpha > 1/2$ shown in the inset of Fig.~\ref{fig_coarsen} and approaches (\ref{rapid2}) for instantaneous quench when $\alpha \to \infty$. For $\alpha < 1/2$, to check if (\ref{slowzs1}) agrees with the numerical results, we need to consider very large $\tau$ and very small $Z$ since the subleading corrections to $G_1(\tau)$ (refer text below (\ref{powCP2})) and to the integral  in (\ref{integ}) (see below (\ref{Heqn})) decay slowly for small $\alpha$. Figure~\ref{fig_coarsen} shows that  in accordance with (\ref{slowzs1}), $(G_1(\tau)-G_1(t))/Z$ diverges as $Z \to 0$ and approaches the prefactor with increasing $\tau$. On comparing  (\ref{rapid2})  with (\ref{slowzs1})-(\ref{slowzs3}), we  arrive at the conclusion that the instantaneous quench model captures the dependence on $Z$ for $\alpha \geq 1/2$ but not otherwise.

\subsubsection{Algebraic cooling} 

For large $\tau$, the equilibrium correlation function ${\cal G}_k(t)=e^{-k \sqrt{2 (1-\gamma)}}$ is equal to one for small $Z$. Comparing (\ref{expCP1}) and (\ref{rapid2}), we find that for any $\beta$, the defect density is exactly given by the instantaneous quench model; this is because of the fact that the algebraic cooling corresponds to $\alpha \to \infty$ in the logarithmic cooling scheme. 

\section{Zero-range process in mean-field geometry}
\label{ZRP}

In this section, we consider a zero-range process \cite{Evans:2005} in mean-field geometry where each site is connected to every other site. 
A site can have $m \geq 0$ identical particles each with mass one, and a particle can hop out to another site with a rate $u(m,t)$ that depends on the number of particles present at the departure site at time $t$. The single-site mass distribution $P(m,t)$ at time $t$ evolves according to the following equations,
\begin{eqnarray}
\frac{dP(0, t)}{dt}&=&u(1, t)P(1, t)-w(t) P(0, t) \label{me0} ~,\\
\frac{dP(m, t)}{dt}&=&u(m+1, t) P(m+1, t)+w(t) P(m-1, t) \nonumber \\
&-&[u(m, t)+w(t)] P(m, t) ~,~m \geq 1~, \label{mem}
\end{eqnarray}
where $w(t)=\sum_{m'=1}^\infty u(m', t)P(m', t)$  is the mean hop rate. Here, we will work with the following time-dependent hop rate, 
\begin{equation}
u(m,t)=1+\frac{b(t)}{m}~,~ m > 0~,
\label{hopp}
\end{equation}   
where 
\begin{equation}
b(t)=b_c \left[1- \left(1-x \right)^\alpha \right]  ~,~x=\frac{t}{\tau}
\label{bcool}
\end{equation} 
that increases from zero to a final value $b_c$ algebraically in time at rate $\tau^{-1}$. In the above equation, $b_c > 2$ (see below) and, as before, the exponent $\alpha > 0$. In this section, we will not consider  the annealing scheme analogous to (\ref{Gexpo}) since it corresponds to the instantaneous quench limit $\alpha \to \infty$, as seen in the last section. We also note that unlike for the correlation function in the Ising model that obeys (\ref{fullGeqn}), the equations of motion for the mass distribution are nonlinear (due to the fugacity term) and the coefficients are also mass-dependent.

When the hop rate (\ref{hopp}) is time-independent, a stationary state exists and exhibits a phase transition from a fluid phase in which particles are homogeneously distributed to a condensate phase where a finite fraction of particles reside in a single mass cluster, as the parameter $b$ is increased keeping the total particle density $\rho_c$  constant \cite{Evans:2005,Priyanka:2016}. The stationary state mass distribution is known exactly to be
\be
{\cal P}(m)=\frac{\omega^m f(m)}{g(\omega)}~,
\label{zrp_ss}
\ee
where $f(m)=\prod_{k=1}^m u^{-1}(k)$ and the partition function, $g(\omega)=\sum_{m=0}^\infty \omega^m f(m)={_2}F_1(1,1; 1+b; \omega)$ is the Gauss hypergeometric function. The total mass density $\rho_c$ is related to the fugacity $\omega=\sum_{m=1}^\infty u(m) {\cal P}(m)$ through the relation $\rho_c=\omega g'(\omega)/g(\omega)$ where prime denotes derivative with respect to $\omega$. For the above $u(m)$, with increasing $b$, the fugacity reaches its maximum value one at a critical value, $b_c=2+\rho_c^{-1}~,~b_c > 2$. Close to the critical point, the steady state distribution ${\cal P}(m) \sim m^{-b} e^{-m/\xi}$ where the static correlation length, $\xi \sim (1-\omega)^{-1} \sim (b_c-b)^{-\nu}$ and the exponent $\nu$ is given by \cite{Priyanka:2014,Priyanka:2016}
\begin{subnumcases}{{\nu} =}
(b_c-2)^{-1} ~&,~$2 < b_c < 3$ \label{nu1}\\
1 ~&,~ $b_c \geq 3$~.
\label{nu2}
\end{subnumcases}
For later reference, we also note that for this model, an auto-correlation function for the mass yields the stationary state dynamical exponent $z=2$ at the critical point  \cite{Godreche:2001}. Furthermore, following an instantaneous quench from the fluid phase to the critical point, the mass distribution ${\hat P}(m,t) = {\cal P}(m,t) {\hat g}_c(m t^{-1/{\hat z}})$ with the critical coarsening exponent ${\hat z}=2$ \cite{Godreche:2003}.

In the following subsections, we study how the mass distribution behaves when the hop rates are time-dependent and given by (\ref{hopp}). 
For convenience, we will work with the ratio $R(m,t)\equiv P(m,t)/{\cal P}(m,t)$ of nonequilibrium to equilibrium probability distribution (instead of the difference between them). 

\subsection{Dynamics deep in the fluid phase}

Our numerical solution of (\ref{me0}) and (\ref{mem}) suggests that the mass distribution $P(m,t)$ has a different scaling form for small and large masses. Therefore, for large $t, \tau$ with $x=t/\tau$ finite, we make the following scaling ansatz,
\bsn
{\frac{P(m,t)}{{\cal P}(m,t)}=}
 1+\tau^{-\theta} \delta P(m,x)~&,~ $m \epsilon \stackrel{\tau \to \infty}{\to 0}$ \label{flu_sm} \\
Q(M,x)~&,~$m \epsilon$ finite ~, \label{flu_lm}
\esn
where the scaling variable $M=m \epsilon(\tau)$ for large $m$ and $\epsilon(\tau)$ decays as a power law with $\tau$. For consistency with (\ref{flu_sm}), we also have $Q(0,x)=1$. Using the above scaling ansatz in the definition of the fugacity, we find that
\bea
w(t,\tau) &=& \int_0^{\epsilon^{-1}} dm {\cal P}(m,x) (1+\tau^{-\theta} \delta P(m,x)) u(m,x)  
\nonumber \\
&+&\epsilon^{b_c-1} \int_1^\infty dM M^{-b_c} e^{-\frac{M}{\epsilon \xi(x)}} Q(M,x) \label{fugsca} \\
&\approx& {\omega(x)} [1+\tau^{-\theta}  \delta w(x)] ~.
\label{flu_fug}
\eea
The second integral on the RHS of (\ref{fugsca}) vanishes in the $\tau \to \infty$ limit as this term is exponentially small in $\epsilon^{-1}$. Thus the  contribution to the fugacity comes from small mass, as one may expect in the fluid phase which is characterized by typical mass of order unity. 

\begin{figure}
\centering
\includegraphics{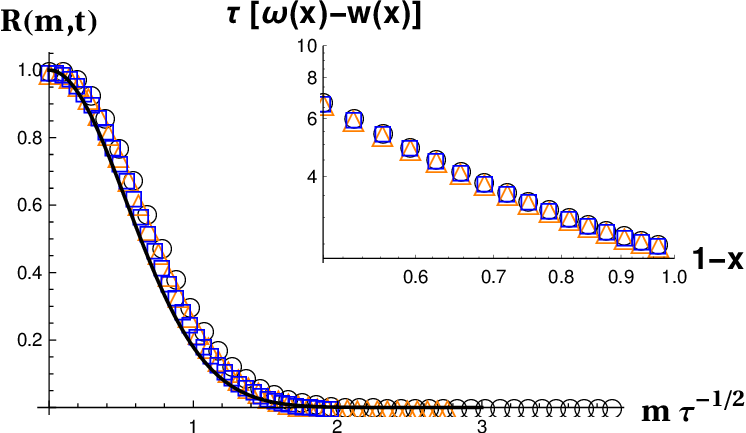}
\caption{Zero-range process: Ratio of the mass distribution function, $R(m,t)=P(m,t)/{\cal P}(m,t)$ for large mass in the fluid phase as a function of the scaled mass $M=m/\sqrt{\tau}$ for the annealing scheme (\ref{bcool})  with $\alpha=1$. The points show the data obtained by numerically solving the master equation (\ref{me0}) and (\ref{mem}) for $x = 1/2$, and the solid line shows the scaling function $Q(M,x)$ given by (\ref{zrp_larm_away}). The  inset shows the data collapse for the scaled fugacity difference $\delta w=\omega-w$ far from the critical point to support the conclusion that the exponent $\theta=1$ in (\ref{flu_fug}). In both the figures, $\tau=2^{16} (\circ), 2^{17} (\vartriangle), 2^{18} (\Box)$ and $b_c = 5/2$.}
\label{fig_zrpf}
\end{figure}

Using (\ref{flu_sm}) and (\ref{flu_fug}) in the master equation (\ref{mem}) for $m \ll \epsilon^{-1}$, we find that the terms on the left-hand side (LHS) of the resulting equation are of the order $\tau^{-1}$ while the RHS is of the order $\tau^{-\theta}$. We therefore deduce that the exponent $\theta=1$, in line with the data collapse shown in the inset of Fig.~\ref{fig_zrpf} for fugacity.  
For large mass, using the scaling ansatz (\ref{flu_lm}) in the master equation (\ref{mem}) and taking the scaling limits $m \to \infty, \epsilon \to 0$ with finite $M$,  we find that the scaling function $Q(M,x)$ obeys the following differential equation, 
\bea
\frac{M Q(M,x)}{\tau \epsilon(\tau)} \frac{d \ln \omega(x)}{d x} 
=\epsilon(\tau)  (\omega(x) -1) \frac{\partial Q(M,x)}{\partial M} ~,
\label{para1}
\eea
where we have also used the exact stationary state distribution (\ref{zrp_ss}). The above equation yields $\epsilon\sim \tau^{-1/2}$ and the scaling function 
\be
Q(M,x)=\exp \left[- \frac{m^2}{2 \tau \omega(x) (1-\omega(x))} \frac{d\omega(x)}{d x} \right]~.
\label{zrp_larm_away}
\ee
Figure~\ref{fig_zrpf} shows that the numerical solution of (\ref{me0}) and (\ref{mem}) for the mass distribution in the fluid phase is in good agreement with (\ref{zrp_larm_away}). 

\subsection{Dynamics close to the critical point} 

For $x \to 1$, we make the following scaling ansatz,
\bsn
{\frac{P(m,t)}{{\cal P}(m,t)}=}
1+\tau^{-\theta} \delta P(m,Z) ~&,~ $m \epsilon \stackrel{\tau \to \infty}{\to} 0$ \label{cr_sm} \\
 Q(M,Z)~&,~$m \epsilon$ finite \label{cr_lm}~,
\esn
where, besides the scaled mass $M=m \epsilon(\tau)$, we have the scaled remaining time $Z=(1-x)\Lambda(\tau)$, $\Lambda(\tau)$ being an algebraically increasing function of $\tau$. Although the correlation length diverges close to the critical point, as we shall see below, $\epsilon \xi$ remains finite but $\tau^{-\theta} \leq \epsilon^{b_c-1}$. As a result, the second term on the RHS of (\ref{fugsca}) for the fugacity vanishes as $\epsilon \to 0$ since $b_c > 2$ (assuming the integral is finite).  It thus follows that for large $\tau$, the fugacity $w(Z)=\omega(x) [1+\tau^{-\theta} \delta w(Z)]$, where the exponent $\theta$ is determined below (see (\ref{theta1}) and (\ref{theta2})).

\begin{figure}
\begin{center}
\includegraphics[width=12cm]{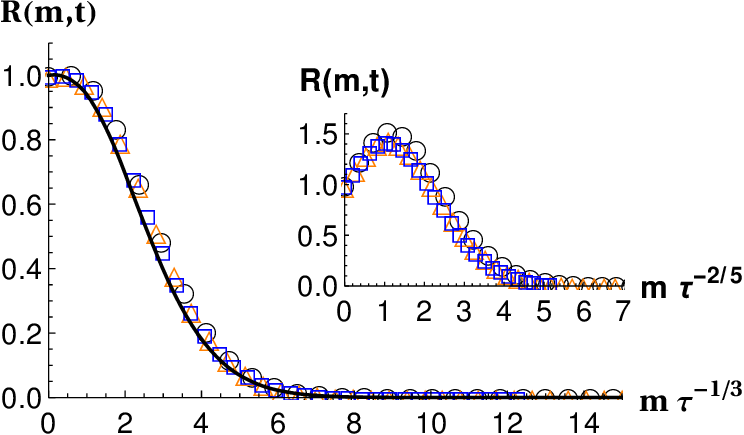}
\caption{Zero-range process: Data collapse for the ratio of the mass distribution functions, $R(m,t)=P(m,t)/{\cal P}(m,t)$ for large mass and close to the critical point to confirm the scaling (\ref{KZzrp1}) and (\ref{KZzrp2})  for the annealing scheme (\ref{bcool})  with $\alpha=1$. The remaining time $Z \approx 0.112$ for $b_c=4$ (main) and $Z \approx 0.051$ for $b_c=5/2$ (inset), and $\tau=2^{17} (\circ), 2^{18} (\vartriangle), 2^{19} (\Box)$. The solid line in the main figure shows the numerical solution of the differential equation (\ref{zrpscafn}) for the scaling function when $b_c > 3$.}
\label{fig_zrpcp}
\end{center}
\end{figure}

For large mass, we use the scaling ansatz (\ref{cr_lm}) in the master equation (\ref{mem}) and find that 
\bea
&&\frac{\epsilon^2}{2} \left(\omega+\frac{w u(m,x)}{\omega} \right)\frac{\partial^2 Q}{\partial M^2}+\epsilon \left(\omega-\frac{w u(m,x)}{\omega} \right)\frac{\partial Q}{\partial M}  \nonumber \\
& +& \left(\omega-u(m,x)-w+\frac{w u(m,x)}{\omega}  \right) Q
= \frac{1}{\tau} \frac{d \ln {\cal P}(m,x)}{dx}Q -\frac{\Lambda}{\tau} \frac{\partial Q}{\partial Z} ~.
\label{scalcomp}
\eea
Using the stationary state properties (\ref{app_zrp1})-(\ref{app_zrp3}) in (\ref{scalcomp}), a simple power counting then yields $\Lambda \sim \tau^{\frac{1}{1+2 \alpha \nu}}$ and $\epsilon \sim \Lambda^{-\alpha \nu}$ so that the scaling variables are given by
\bea
M=m \tau^{-\frac{\alpha}{b_c+2 \alpha-2}} ~&,~ Z= (1-x) \tau^{\frac{b_c-2}{b_c+2 \alpha-2}}~&,~ 2< b_c < 3 \label{KZzrp1}\\
M= m \tau^{-\frac{\alpha}{1+2 \alpha}}~&,~Z=(1-x) \tau^{\frac{1}{1+2 \alpha}}~&,~b_c \ge 3 \label{KZzrp2}
\eea
The above scaling laws are consistent with the Kibble-Zurek argument which predicts that $\xi(x_*) \sim \tau^{\frac{\alpha \nu}{1+\alpha \nu z}}, 1-x_* \sim \tau^{-\frac{1}{1+\alpha \nu z}}$ (see also Fig.~\ref{fig_zrpcp}). 

The above analysis also shows that the scaling function $Q(M,Z)$ obeys the following differential equation, 
\be
\frac{\partial^2 Q}{\partial M^2}-\left(C Z^{\alpha \nu}+\frac{b_c}{M}\right) \frac{\partial Q}{\partial M} +\frac{\partial Q}{\partial Z} - \alpha \nu C M Z^{\alpha \nu-1} Q 
= \frac{\tau^{-\theta} \delta w}{\epsilon}   \left(\frac{\partial Q}{\partial M} - C Z^{\alpha \nu} Q-\frac{b_c}{M} Q\right)
\label{zrpscafn}
\ee
where the constant $C$ is given by (\ref{app_Ceqn1}) and (\ref{app_Ceqn2}). The $\delta w$-term on the RHS of the above differential equation contributes if $\tau^{-\theta}  \sim \epsilon $ which, as shown below in (\ref{theta1}), is true for $b_c < 3$. Thus, on setting the RHS of (\ref{zrpscafn}) to zero for $b_c > 3$, we obtain a closed equation for $Q(M,Z)$ whose numerical solution subject to boundary conditions $Q(0,Z)=1, Q(\infty, Z)=0, Q(M,\infty)=0$ matches well with that obtained using the master equation as shown in Fig.~\ref{fig_zrpcp}.

\begin{figure}
\begin{center}
\includegraphics[width=12cm]{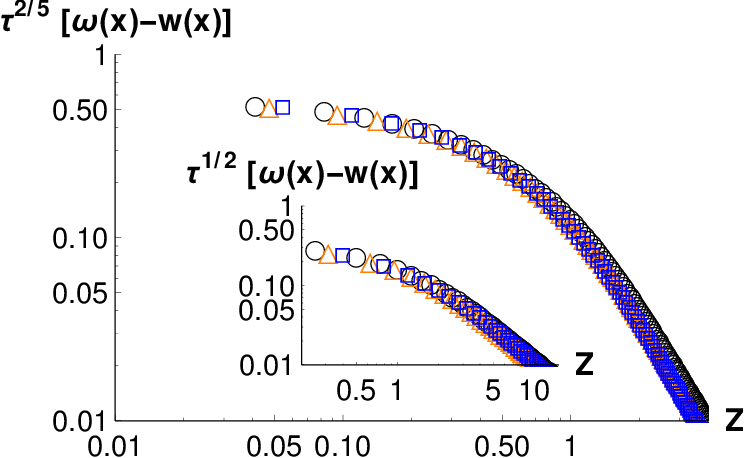}
\caption{Zero-range process: Data collapse for the scaled fugacity difference, $\delta w (Z)=\tau^{\theta} (w-\omega) $ for the annealing scheme (\ref{bcool})  with $\alpha=1$, $b_c=5/2$ (main) and $7/2$ (inset), and $\tau=2^{17} (\circ), 2^{18} (\vartriangle), 2^{19} (\Box)$ to verify the $\theta$-exponent given by (\ref{theta1}) and (\ref{theta2}).}
\label{fig_fug}
\end{center}
\end{figure}

To find the exponent $\theta$, we adapt the analysis of \cite{Godreche:2003} for the coarsening dynamics of zero-range process with time-independent hop rates to the model considered here. Using (\ref{cr_sm}) in the master equation (\ref{mem}) for small mass, we arrive at
\be
\frac{\tau^\theta}{\tau} \frac{d \ln {\cal P}(m,x)}{dx} \approx \Delta P(m+1,Z)-u(m,1) \Delta P(m,Z) +\delta w(x) (u(m,1)-1) 
\label{thetam} ~,
\ee
where $\Delta P(m,Z)=\delta P(m,Z)-\delta P(m-1,Z)$. 
On applying (\ref{thetam}) to small enough mass so that the LHS vanishes for large $\tau$, we find that $\delta P(m,Z)$ increases linearly with $m$. Using this result in the mass conservation equation yields the fugacity exponent,
\bsn
{\theta =}
\frac{\alpha}{b_c-2+2 \alpha} ~&,~ $2 < b_c < 3$ \label{theta1}\\
\frac{\alpha (b_c-2)}{1+2 \alpha } ~&,~ $b_c \geq 3$ \label{theta2}~. 
\esn
We verify that in the $\alpha \to \infty$ limit, (\ref{theta2}) reduces to the result (27) of \cite{Godreche:2003} for the exponent ${\hat \theta}$ after an instantaneous quench to the critical point; moreover, the above equation predicts the corresponding exponent for $2 < b_c < 3$ to be $1/2$. The exponent $\theta$ is numerically tested in Fig.~\ref{fig_fug}, and we find a good agreement with (\ref{theta1}) and (\ref{theta2}). 

\begin{figure}
\begin{center}
\includegraphics[width=12cm]{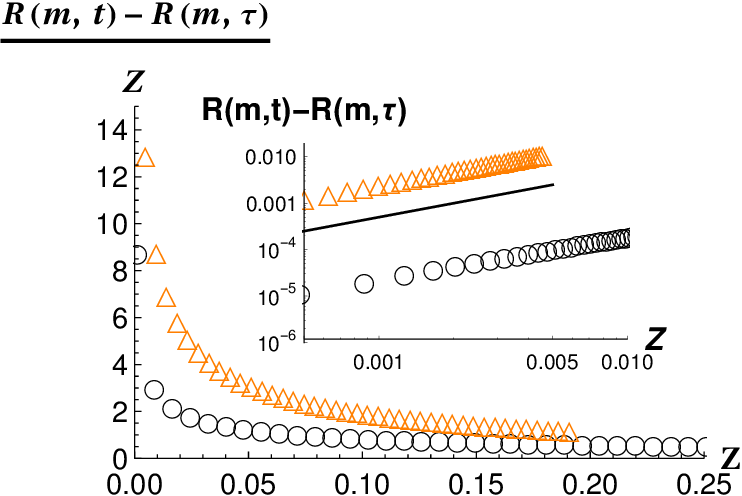}
\caption{Zero-range process: Variation of the ratio of the mass distribution, $R(m,t)=P(m,t)/{\cal P}(m,t)$ close to the critical point with the scaled remaining time $Z$ for $M=2, b_c=5/2 (\vartriangle)$ and $M=4, b_c=7/2 (\circ)$ with $\tau=2^{20}$ for annealing scheme (\ref{bcool}). The difference $R(m,t)-R(m,\tau)$ decays sublinearly with $Z$ for $\alpha=1/2$ (main),  and linearly for $\alpha=1$ (inset) as indicated by solid line and in agreement with the instantaneous quench model prediction (\ref{zrpco}).}
\label{fig_zcoarse}
\end{center}
\end{figure}

\subsection{Relation to the coarsening process}

The fugacity exponent $\theta$ found above may be understood on invoking critical coarsening close to the critical point. At the end of the quench, the deviation in fugacity and distribution of small mass are expected to decay with the time remaining until the end of the quench as a power law with an exponent ${\hat \theta}$ \cite{Biroli:2010}. Using (\ref{KZzrp1}) and (\ref{KZzrp2}) and the results of \cite{Godreche:2003} for ${\hat \theta}$, we find that (\ref{theta1}) and (\ref{theta2}) are indeed consistent with this expectation. The deviation in the small mass distribution at the end of the quench has also been studied in the one-dimensional version of this model \cite{Priyanka:2016}. Specifically, the scaling of the difference $P(0,\tau)-{\cal P}(0,\tau)$ with the annealing rate was studied as it is related to the excess defect density in the one-dimensional exclusion process, and it was found numerically that the mass deviation scales as $(\tau-t_*)^{-{\hat \theta}}$ on using the exponents in the one-dimensional zero-range process \cite{Priyanka:2016}.

To understand the $Z$-scaling of the mass distribution, using (28) of \cite{Godreche:2003} for an instantaneous quench to the critical point, we first note that for large mass, 
\bea
\frac{{\hat {P}}(m,t)}{{\cal P}(m,t)} &=& 1-\frac{1}{2^{1+b_c} \Gamma(\frac{3+b_c}{2})} \left(\frac{m}{\sqrt{t-t_*}} \right)^{1+b_c} \\
&=& 1-\frac{1}{2^{1+b_c} \Gamma(\frac{3+b_c}{2})} \left(\frac{M}{\sqrt{1-Z}} \right)^{1+b_c} \\
&\approx& 1-\frac{M^{1+b_c}}{2^{1+b_c} \Gamma(\frac{3+b_c}{2})} -\frac{M^{1+b_c} Z}{2^{1+b_c} \Gamma(\frac{1+b_c}{2})} ~,
\label{zrpco}
\eea
where ${\hat {P}}(m,t)$ denotes the mass distribution on instantaneous quench. 
Thus, close to the critical point, the ratio of the mass distribution decays linearly with the scaled remaining time $Z$, when the system is quenched infinitely fast. 

To see if this linear behavior holds when quenching occurs at a finite rate, we studied the ratio $R(m,t)=P(m,t)/{\cal P}(m,t)$ numerically. As our results displayed in Fig.~\ref{fig_zcoarse} show, $R(m,t)$ decays linearly with $Z$ for $\alpha=1$ but not for $\alpha=1/2$. Although we are not able to show it analytically, our numerical results suggest that the linear scaling holds for $\alpha \geq 1$ and $R(m,t)-R(m,\tau)$ varies as $Z^\alpha$ for $\alpha \leq 1$. We have also studied how the mass variance changes with $Z$ and find that it also varies sublinearly for sufficiently small $\alpha$ (data not shown).

\section{Discussion}
\label{disc}

In this article, we have analyzed in detail the nonequilibrium dynamics of a classical system when it is annealed slowly from a disordered phase to the critical point in the framework of a kinetic Ising model and a zero-range process. The Kibble-Zurek argument that explains how the equilibrium is approached with decreasing annealing rate has been verified numerically in various recent studies \cite{Jelic:2011,Liu:2014,Priyanka:2016,Ricateau:2018,Jeong:2019,Jeong:2020}. But it has also been found that this argument overestimates the defect density at the end of the quench, and scaling laws different from those predicted by it can be obtained when critical and subcritical coarsening are taken into account \cite{Biroli:2010}. However, the previous body of work does not give any insight into the associated scaling functions. 

For the Ising model, we have derived the scaling function using the exact solution (\ref{Ispl}) (see also Appendix~\ref{app_ising}) for the spin-spin correlation function. For the zero-range process, the scaling function is found to obey (\ref{zrpscafn}) which, however, does not appear to be solvable and was studied numerically.
These results shed light on the role of critical coarsening in the slow quench dynamics: 
working with the annealing scheme (\ref{Gpower}) and (\ref{bcool}), we find that close to the critical point, $V(K,Z)$ which represents domain wall density in the Ising model and nonequilibrium mass distribution (relative to the equilibrium one) in the zero-range process is of the following form, 
\bsn
{{V}(K, Z)=}
A_1(K, \alpha)+A_2(K, \alpha) Z^{a}~,&~ $\alpha < \alpha_c$ \\
{\hat A}_1(K, \alpha)+{\hat A}_2(K, \alpha) Z~,&~ $\alpha > \alpha_c$~,
\esn
where $K$ and $Z$ are the Kibble-Zurek length and time scales, respectively.  In the above equation, the prefactors are nonuniversal in that they depend on the details of the cooling protocol. But if the quench exponent $\alpha > \alpha_c$, the scaling with $Z$ is found to be linear. 
This behavior should be compared with the corresponding result following an instantaneous quench where ${\hat V}(K,Z)={\hat V}_1(K)+{\hat V}_2(K)Z$ for small $Z$; we thus conclude that the instantaneous quench model can describe the slow quench dynamics if the annealing occurs sufficiently fast. The interpretation and physical origin of the sublinear scaling is however not clear.

We close this article with some open questions and directions. Here we have quenched the system to the critical point, but a nontrivial critical point $b_c$ exists in the zero-range process. We have performed a preliminary numerical study to understand the scaling behavior when the system is driven across the critical point, and find that the nonequilibrium mass distribution, 
$P(m,\tau)=\tau^{-1} f(m/\sqrt{\tau})$ for large mass which is consistent with the expectation that $m \sim \tau^{1/{\hat z_O}}$ where ${\hat z_O}=2$ is the coarsening exponent in the mean-field zero range process when the system is quenched rapidly to the condensate phase \cite{Godreche:2003}; an analogous result has also been obtained for the one-dimensional zero-range process \cite{Priyanka:2016}. However, a more detailed analysis is needed to understand the dynamics close to the final quench point in the condensate phase.
It would be instructive to study the slow quench dynamics for the Ising chain with Kawasaki dynamics where, unlike for Glauber dynamics studied here, the dynamic exponent $z$ and the coarsening exponent ${\hat z}$ are not equal, and one may expect the defect density at the end of the quench to exhibit a scaling different from that predicted by Kibble-Zurek argument \cite{Biroli:2010}. More general schemes such as series of cooling-heating cycles that in the context of glasses \cite{Ritort:2003} are known to show hysteretic effect could also be interesting to study. Most of the previous work on high-dimensional models have assumed linear annealing \cite{Biroli:2010,Priyanka:2016}; it would be interesting to explore the dependence on other functional forms and determine the critical quench exponent above which instantaneous quench model works. An analytical understanding of the mean-field zero-range process presented here and the one-dimensional zero-range process in \cite{Priyanka:2016}, and more generally, models with nontrivial critical point remains an open question for future work.


\clearpage
\appendix
\setcounter{equation}{0}
\renewcommand{\thesection}{\Alph{section}}
\numberwithin{equation}{section}


\section{Zero-range process: stationary state properties}
\label{app_ss}

From the density-fugacity relation $\rho=\omega g'(\omega)/g(\omega)$, we have
\be
\rho_c=\frac{1}{b_c-2}= \frac{\omega}{b+1} \frac{{_2}F_1(2,2; 2+b; \omega)}{{_2}F_1(1,1; 1+b; \omega)}~.
\ee
Using (15.3.6) of \cite{Abramowitz:1964}, we expand the RHS of the above equation about $\omega=1$. For $b > 3$, we get
\be
\frac{1}{b_c-2} \approx \frac{1}{b-2} \left(1- \frac{(b-1)^2 (1-\omega)}{(b-2)(b-3)} \right)
\ee
Similarly, for $2 < b < 3$, we have
\be
\frac{1}{b_c-2} \approx \frac{1}{b-2}- \frac{\pi (b-1)^2}{\sin(\pi b)} (1-\omega)^{b-2}
\ee
Using the annealing scheme (\ref{bcool}) when $x \to 1$ then leads to 
\bsn
{C(b_c)=}
\left(\frac{b_c \sin(\pi b_c)}{\pi (b_c-2)^2 (b_c-1)^2} \right)^{\frac{1}{b_c-2}}~&,~   $2 < b_c < 3$ \label{app_Ceqn1}\\
\frac{b_c (b_c-3)}{(b_c-1)^2}~&,~   $b_c > 3$ \label{app_Ceqn2}~.
\esn

From the exact stationary state solution (\ref{zrp_ss}), close to the critical point, we also obtain
\bea
\omega-u(m,x) &\approx& \omega-1-\frac{b_c}{m}  \label{app_zrp1}\\
1-\omega(x) &\approx& C(b_c)  (1-x)^{\alpha \nu} \\
\frac{d \ln {\cal P}(m,x)}{dx} &\approx& (m-\rho_c) C \alpha \nu (1-x)^{\alpha \nu-1}\nonumber \\
&+& \alpha b_c (\psi(b_c+1)-\psi(b_c+m+1)) (1-x)^{\alpha-1} ~, \label{app_zrp3}
\eea
where $\psi(n)$ is the digamma function \cite{Abramowitz:1964}.

\section{Kinetic Ising model: scaling function}
\label{app_ising}

In the main text, an exact solution is obtained for the correlation function $G_k(t)$ which is defined in discrete space. But using the scaling variable $K$, one can go from discrete to continuous space and obtain a partial differential equation for the scaling function $F(K,Z)$  at low temperatures. Working with the scaling variables $K$ and $Z$ defined in (\ref{KZexpo1}) in the exact equation (\ref{fullGeqn}), we obtain 
\be
\frac{\partial^2 F}{\partial K^2}+\frac{\partial F}{\partial Z}-2 Z^\alpha F + \frac{\alpha}{\sqrt{2}} K Z^{\frac{\alpha}{2}-1} {\cal G}(K,Z)=0~.
\label{app_pde2}
\ee 
with $F(0,Z)=F(\infty,Z)=F(K,\infty)=0$. 
Taking a sine transform on both sides of the above equation with respect to $K$, we obtain a first order differential equation in $Z$ which can be easily solved. The inverse sine transform then yields (\ref{gpowCk}) in the main text. 


\clearpage

\bibliography{/Users/justin/Desktop/SlowQuench/SM_papers/masterKZ.bib}
\bibliographystyle{iopart-num}


\end{document}